\documentclass[prb,aps,twocolumn]{revtex4}
\usepackage{graphicx}% Include figure files
\usepackage{epstopdf}
\usepackage{dcolumn}% Align table columns on decimal point
\usepackage{bm}% bold math
\usepackage{color}
\usepackage{ulem}
\usepackage{amsfonts,amsthm}
\usepackage{amsmath}
\usepackage{amssymb}
\usepackage{braket}
\begin{document}
\newcommand{\bR}{\mbox{\boldmath $R$}}
\newcommand{\tm}[1]{\textcolor{red}{#1}}
\newcommand{\tb}[1]{\textcolor{blue}{#1}}
\newcommand{\tbs}[1]{\textcolor{blue}{\sout{#1}}}
\newcommand{\Ha}{\mathcal{H}}
\newcommand{\mh}{\mathsf{h}}
\newcommand{\mA}{\mathsf{A}}
\newcommand{\mB}{\mathsf{B}}
\newcommand{\mC}{\mathsf{C}}
\newcommand{\mS}{\mathsf{S}}
\newcommand{\mU}{\mathsf{U}}
\newcommand{\mX}{\mathsf{X}}
\newcommand{\sP}{\mathcal{P}}
\newcommand{\sL}{\mathcal{L}}
\newcommand{\sO}{\mathcal{O}}
\newcommand{\la}{\langle}
\newcommand{\ra}{\rangle}
\newcommand{\ga}{\alpha}
\newcommand{\gb}{\beta}
\newcommand{\gc}{\gamma}
\newcommand{\gs}{\sigma}
\newcommand{\vk}{{\bm{k}}}
\newcommand{\vq}{{\bm{q}}}
\newcommand{\vR}{{\bm{R}}}
\newcommand{\vQ}{{\bm{Q}}}
\newcommand{\vga}{{\bm{\alpha}}}
\newcommand{\vgc}{{\bm{\gamma}}}
\newcommand{\mb}[1]{\mathbf{#1}}
\def\vec#1{\boldsymbol #1}
\arraycolsep=0.0em
\newcommand{\Ns}{N_{\text{s}}}
%
%\preprint{APS/123-QED}

\title{
{Real-time evolution and quantized charge pumping in magnetic Weyl semimetals }
}

\author{
Takahiro Misawa$^1$, Ryota Nakai$^2$ and Kentaro Nomura$^{2,3}$
}

\affiliation{$^1$Institute for Solid State Physics,~University of Tokyo,~5-1-5 Kashiwanoha, Kashiwa, Chiba 277-8581, Japan}
\affiliation{$^2$Institute for Materials Research,~Tohoku University,~Sendai 980-8577,~Japan}
\affiliation{{$^3$Center for Spintronics Research Network, Tohoku University, Sendai, 980-8577, Japan}}

\date{\today}

\begin{abstract}
{Real-time evolution and charge pumping} in {magnetic Weyl semimetals are} studied
by solving the time-dependent Schr\"{o}dinger equations.
In the adiabatic limit of {the real-time evolution,} 
we show that the total pumped charge is quantized in the {magnetic Weyl semimetals}
as in the quantum Hall system although the Weyl semimetal has no bulk gap.
We examine how the disorder 
affects  
the {charge} pumping.
As a result, we show that the quantized pumped charge 
is robust against the small disorder and
find that the pumped charge increases
in the intermediate disorder region.
We also examine the doping effects on the 
{charge} pumping and
show that the remnant of the quantized pumped charge at zero doping 
can be detected.
Our results show that the 
{real-time evolution} is a
useful technique for detecting the topological 
properties of the systems with 
no bulk gap and/or disorders.
\end{abstract}

\pacs{to be determined}

\maketitle
%----- Introduction -----

\section{Introduction}
\begin{figure}[b!]
  \begin{center}
    \includegraphics[width=7cm,clip]{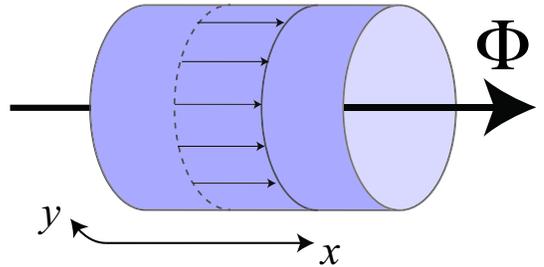}
  \end{center}
\caption{(color online)~Schematic picture 
for the Laughlin's gedanken experiment in the quantum Hall system.
By introducing flux $\Phi$ 
(equivalently introducing electric field in $y$ direction), 
charge pumping occurs in $x$ direction.}
\label{fig:cylinder}
\end{figure}

Immediately after the discovery of the integer quantum Hall effects~\cite{Klitzing_PRL1980}, 
Laughlin presented the simple and important gedanken experiment for explaining the
quantized Hall conductivity~\cite{Laughlin_PRB1981}. 
In the Laughlin's gedanken experiment, 
by adiabatically introducing the magnetic flux $\Phi$ from 0
to $\Phi=\Phi_{0}=e/h$ into 
the quantum Hall system on a cylinder, 
the electrons move from one edge to opposite side of edge 
as schematically shown in Fig.~\ref{fig:cylinder}.
Due to the invariance of the wave functions 
under the gauge transformation by the flux $\Phi$,
it is shown that the
total pumped charge should be quantized when $\Phi=\Phi_{0}$.

In a similar way,
Thouless argued the charge pumping in the one-dimensional systems with 
the slow time-dependent periodic potentials~\cite{Thouless_PRB1983}.
By solving the time-dependent Schr\"{o}dinger equations, 
Thouless shows that the pumped charge is quantized  
and it is related with the topological invariant.
The 
charge pumping caused by introducing the external flux
is called {the} {\it Thouless pumping}
and {the} Laughlin's argument can be regarded as 
the adiabatic limit of the Thouless pumping. 
Although the realization of the 
Thouless pumping in experiment is difficult because
introducing the magnetic flux or {adiabatically} 
controlling the periodic potential are difficult,
recent experiments show that {the} Thouless pumping 
can be realized in the 
ultra cold atoms~\cite{Nakajima_NPhys2016,Lohse_NPhys2016}.

{In a theoretical point of view,
{the} Thouless pumping is a useful theoretical technique}
for detecting the topological invariant.
In the previous studies,
the Thouless pumping in the
quantum Hall system is numerically studied and
it is shown that the charge pumping continuously occurs from
$t=0$ ($t$ represents time) and it reaches the quantized value 
at $t=T$ ($T$ is the time interval during which the magnetization increases by $\Phi_{0}$) 
in the adiabatic limit~\cite{Maruyama_JPC2009,Hatsugai_PRB2016}.
{In} the quantum Hall systems,
the total charge pumping is expressed by the 
topological invariant as follows~\cite{Thouless_PRL1982,Kohmoto_AP1985}:
\begin{align}
\Delta N(t=T)=N_{\rm L}(t=T)-N_{\rm R}(t=T)= 2\times \mathcal{C},
\end{align}
where $N_{\rm L}(t)$ ($N_{\rm R}(t)$) 
{denotes the number of electrons distributed in left side (right side) of the system %, %at time $t$,
and $\mathcal{C}$ is the topological invariant called the Chern number 
that takes integers ($\mathcal{C}=0,\pm 1,\pm2, \cdots$).

In the conventional ways for calculating the topological invariants, 
it is necessary to define the Bloch wave functions{~\cite{Ryu2010}}.
Although such definitions are useful for non-interacting systems with 
{translational invariance,}
they are not directly used for non-periodic systems such as the disordered systems.
In the Thouless pumping, by solving the
time-dependent Schr\"{o}dinger equations,
it is easy to calculate the topological invariant even for disordered systems 
through the quantized charge pumping.
We note that Thouless pumping may be useful for detecting the
topological invariant in the correlated electron systems{~\cite{Nakagawa2018PRB}}.

\begin{figure}[t!]
  \begin{center}
    \includegraphics[width=8cm,clip]{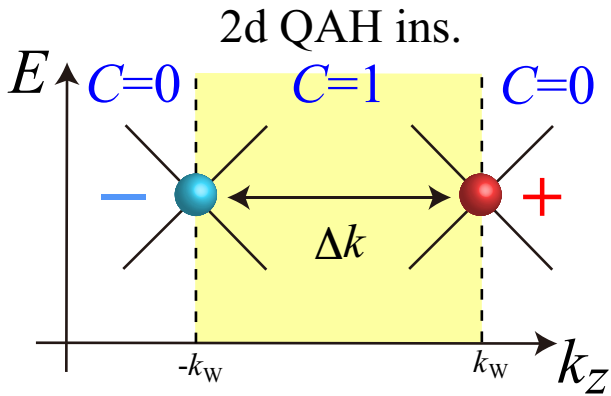}
  \end{center}
\caption{
(color online)~Schematic picture of the 
Weyl points in the momentum space.
The positions of the Weyl points are denoted by $\pm k_{\rm W}$.
Signs of monopole charges~\cite{Hirayama_PRL2015} 
at the Weyl points are represented by $+$ and $-$.
Inside the Weyl point ($-k_{\rm W}\leq k_{z}\leq k_{\rm W}$), 
the Chern number is non trivial ($\mathcal{C}=1$)
while it is trivial outside the Weyl points ($\mathcal{C}=0$).
The Hall conductivity in the Weyl semimetal is given by
{$\sigma_{xy}=\frac{e^2}{h}\frac{\Delta k}{2\pi}$}.}
\label{fig:Weyl}
\end{figure}

{In this paper, {by using the real-time evolution}, 
we apply the Thouless pumping
to the Weyl semimetals where the quantized charge 
pumping also occurs~\cite{Shindou_PRL2001,Murakami_NJ2007,Wan_PRB2011,Burkov_PRL2011,Burkov2018}.}
We note that, Weyl semimetals have been recently 
found in inversion symmetry broken systems such as
TaAs~\cite{Xu_Scicenc2015,Lv_PRX2015,Lv_NPhys2015,Yang_NPhys2015} and
time-reversal symmetry broken systems (magnetic Weyl semimetals) such as 
Mn$_{3}$Sn~{\cite{Kuroda_NPhys2017,Ito2017JPSJ}},
Heusller alloys~\cite{Wang2016PRL,Chang2016SR}, 
Co$_3$Sn$_2$S$_2$~{\cite{Liu2018,Xu2018,Wang2018,Ozawa2019}},
and Sr$_{1-y}$Mn$_{1-z}$Sb$_{2}$~\cite{Liu2017}.
Because the magnetic Weyl semimetals can be constructed by stacking the
two-dimensional quantum anomalous Hall (QAH) systems (see Fig.2),
it is shown that the Hall conductivity is quantized  as follows:
\begin{align}
\sigma_{xy}&=\frac{e^2}{h}\frac{\Delta k}{2\pi}, 
\end{align}
where $\Delta k$ is the distance between two Weyl points.
By performing the Thouless pumping in the Weyl semimetal,
it is expected that the charge pumping is also quantized as follows:
\begin{align}
\Delta N &= 2\Delta k \times\frac{L_{z}}{2\pi},
\end{align}
where  $L_{z}$ is the length of the {system} in $z$ direction.
In the Weyl semimetals, however, the charge gap is zero 
and it is non-trivial whether the Thouless pumping gives the quantized
charge pumping for gapless systems or not.
{In this work,} we show 
that the 
quantized charge pumping
occurs in the adiabatic limit.
This result indicates that the Thouless pumping is useful
even when the bulk charge gap is absent.

We also examine the effects of the disorder on the Thouless
pumping and find that the 
{charge pumping} is robust against
the small disorder and it increases in the intermediate
disorder region. These behaviors are consistent with 
the previous studies~\cite{Chen_PRL2015,Shapourian_PRB2016,Takane_JPSJ2016}.
This 
indicates that the Thouless pumping {also} works well
for disordered systems.

This paper is organized as follows:
In Sec.2.A, we introduce model Hamiltonians for 
describing the Weyl semimetal and
explain the algorithms for solving the
time-dependent Schr\"{o}dinger equation in Sec.2.B.
Although the algorithms 
are explained in the literature~{\cite{Suzuki1990PhyLettA,Suzuki1993PJA,suzuki1994,nakanishi1997}},
to make our paper self contained,
we detail how to efficiently solve the time-dependent Schr\"{o}dinger equations. 
In Section 3.A, we show the results of the Thouless pumping
for clean limit and at zero doping. 
Then, we examine the disorder effects in Sec. 3.B.
We also examine the doping effects in Sec.3.C and
show that the Thouless pumping {occurs} %works well 
for the 
finite doping case, i.e., remnant of the quantization can be detected.
Finally, Section 4 is devoted to the summary. 

\section{Model and Method}
\subsection{Lattice Model for Weyl semimetals}
The Hamiltonian used in this study is given by  
\begin{align}
&H_{\rm W}=\sum_{\nu=x,y,z}H_{\nu}+H_{\rm diag},\\
&H_{\nu}=\sum_{j}h_{\nu,j},\\
&{h_{\nu,j}=c_{j+\vec{e}_{\nu}}^{\dagger}T_{\nu}c_{j}+{\rm H.c.}},\\
&H_{\rm diag}=(2{t_{\rm hop}}-m)\sum_{j}c_{j}^{\dagger}\sigma_{z}c_{j}+\sum_{j}\epsilon_{j}c_{j}^{\dagger}\sigma_{0}c_{j},\\
&\epsilon_{j}\in [-W/2,W/2].
\label{eq:Weyl}
\end{align}
where $c_{j}^{\dagger}$ ($c_{j}$)
represents the two-component fermion creation (annihilation)
operator defined on a site $j$ on the three dimensional cubic lattice 
spanned by three orthogonal unit vectors $\vec{e}_{\nu=x,y,z}$.
The matrices $T_{\nu}$ are defined as
\begin{align}
T_{x}&={t_{\rm hop}}(-\sigma_{z}+{\rm i}\sigma_{x})/2,\\
T_{y}&={t_{\rm hop}}(-\sigma_{z}+{\rm i}\sigma_{y})/2,\\
T_{z}&=-{t_{\rm hop}}\sigma_{z}/2,
\end{align}
where 
\begin{align}
\sigma_{x}=
\begin{pmatrix}
0 &~1 \\
1 &~0 \\
\end{pmatrix},
\sigma_{y}=
\begin{pmatrix}
0       &~-{\rm i} \\
{\rm i} &~0 \\
\end{pmatrix},
\sigma_{z}=
\begin{pmatrix}
1  &~0 \\
0  &~-1 \\
\end{pmatrix}.
\end{align}
{Throughout this paper, we take the amplitude of 
the hopping transfer $t_{\rm hop}=1$ as a unit of the energy scale.}

The band structure of the Hamiltonian is given by
\begin{align}
&{E_{\rm W}}={\pm}
[(\sin{k_{x}})^2+(\sin{k_{y}})^2 \notag \\
&+(2-m-\cos{k_x}-\cos{k_y}-\cos{k_z})^2]^{1/2}
\end{align}
From this band structure, 
we can show that the Weyl points are located at 
{$(0,0,k_{\rm W}=\pm\cos^{-1}(-m))$}
for $|m|<1$.
Around the Weyl points, the dispersions are given as
\begin{align}
&{E_{\rm W}}\sim\pm (k_{x}^{2}+k_{y}^{2}+\tilde{k}_{z}^{2})^{1/2},
\end{align}
where {$\tilde{k}_{z}=\sqrt{1-m^2}(k_{z}-\cos^{-1}(-m))$}.
We note that 
the Weyl semimetal is constructed by 
stacking the two dimensional QAH insulators.
As shown in Fig.~\ref{fig:Weyl},
the Chern number becomes nontrivial inside {between}
the {two} Weyl points and it becomes trivial outside the 
Weyl points.
Thus, the quantized Hall conductivity is proportional to $\Delta k=2k_{\rm W}$.

To perform the Thouless pumping, we introduce the 
time-dependent vector potentials as follows:
\begin{align}
&T_{y}(t)=e^{{\rm i}A_{y}(t)}\times T_{y},\\
&A_{y}(t)=\frac{2\pi t}{L_{y}T}. \label{eq:vecp}
\end{align}
By introducing $A_{y}(t)$ in $y$ direction,
the charge pumping in $x$ directions occurs
if the Hall conductivity is finite.

\subsection{Method for solving the time-dependent Schr\"{o}dinger equations}
To perform the Thouless pumping,
we explicitly solve the time-dependent  Schr\"{o}dinger equation defined as 
\begin{align}
{\rm i}\frac{\partial\ket{\phi(t)}}{\partial t}=H_{\rm W}\ket{\phi(t)}.
\end{align}
Here, $\ket{\phi(t)}$ is a single Slater determinant given by
\begin{align}
\ket{\phi(t)}=\prod_{n=1}^{N_{\rm e}}\Big(\sum_{i=0}^{N_{\rm s}-1}
\Phi_{ni}(t){c}_{i}^{\dagger}\Big)\ket{0},
\end{align}
where $N_{\rm e}$ is number of particles,
$N_{\rm s}$ is number of sites, 
{and $\Phi_{ni}(t)$ denotes the coefficient of the Slater determinant.}
By discretizing the time and multiplying the time-evolution operator
$U(t+\Delta t,t)$ to the wave functions at each discretized time step,
we can solve the time-dependent Schr\"{o}dinger equations as follows: 
\begin{align}
\ket{\phi(t+\Delta t)}&=U(t+\Delta t,t)\ket{\phi(t)}, \\
U(t+\Delta t,t)&=\mathcal{T}\Big(\exp{\Big[-{\rm i}\int_{t}^{t+\Delta t}H_{W}(s){\rm d}s]}\Big),
\end{align}
{where $\mathcal{T}$ it the time ordering operator.}

One simple way to solve the time-dependent  Schr\"{o}dinger
equation is given by
\begin{align}
|\phi(t+\Delta t)\rangle\sim\exp{(-{\rm i}\Delta tH_{W}(t))}|\phi(t)\rangle.
\end{align}
Here, we approximate $U(t+\Delta,t)$ as
$\exp{(-{\rm i}\Delta tH_{W}(t))}$
{and we denote the coefficients of the
Hamiltonian as $\tilde{H}_{W}$, i.e., 
$H_{W}(t)={c}^{\dagger}\tilde{H}_{W}(t){c}$.}
By diagonalizing $\tilde{H}_{W}(t)$ at each time step,
we obtain the solutions as follows:
\begin{align}
{e^{-{\rm i}\Delta t\cdot {c}^{\dagger}\tilde{H}_{W}(t){c}}\ket{\phi(t)}}
&=\prod_{n=1}^{N_{\rm e}}\Big(\sum_{i=0}^{N_{\rm s}-1}
{\Phi}_{ni}(t+\Delta t){c}_{i}^{\dagger}\Big)\ket{0},\\
{\Phi}_{ni}(t+\Delta t)&=
{\sum_{\alpha j}{\Phi_{n j}(t)}e^{-{\rm i}\Delta t\lambda_{\alpha}}(V^{\dagger})_{\alpha j}V_{i\alpha}},\\
V^{\dagger}{\tilde{H}_{W}(t)}V&={\rm diag}(\lambda_{0},\lambda_{1},\dots,{\lambda_{2N_{\rm s}-1}}).
\end{align}
Because this method requires the diagonalization of the 
Hamiltonian at each step, the computational costs are large.
To reduce the costs,
we decompose the time-evolution operator by using the 
Suzuki-Trotter decomposition~{\cite{Suzuki1990PhyLettA,Suzuki1993PJA,suzuki1994}}.
In this method, because the diagonalization of the 
full Hamiltonian is necessary only for preparing the initial
wave functions, computational cost is drastically reduced.

From here, 
we explain outline of the method.
{Since the cubic lattice is bipartite, we decompose the nearest-neighbor-hopping 
terms in the Hamiltonian into two parts as follows:}
\begin{align}
{H}_{\nu}&={H}_{\nu,e}+{H}_{\nu,o},\\
{H}_{\nu,e}&={h}_{\nu,0}+{h}_{\nu,2}+\cdots,\\
{H}_{\nu,o}&={h}_{\nu,1}+{h}_{\nu,3}+\cdots,
\end{align}
{where $h_{\nu,2n~(2n+1)}$ 
contains hopping terms between sites on 
$\nu=2n~(2n+1)$ and those on $\nu=2n+1~(2n+2)$ for $\nu=x,y,z$.}
We note that each component of the Hamiltonian can be
described as
\begin{align}
h_{\nu,i}=
\begin{pmatrix}
{c}_{i}^{\dagger} &~{c}_{i+{\vec{e}_{\nu}}}^{\dagger}
\end{pmatrix}
K_{\nu}
\begin{pmatrix}
{c}_{i} \\
{c}_{i+{\vec{e}_{\nu}}}
\end{pmatrix}.
\end{align}
For example, $K_{x}$ is given by
\begin{align}
K_{x}=\frac{1}{2}
\begin{pmatrix}
0 & ~0 & ~-1 & ~-{\rm i} \\ 
0 & ~0 & ~-{\rm i} & ~1 \\ 
-1 & ~{\rm i} & ~0 & ~0 \\ 
{\rm i} & ~1 & ~0 & ~0 \\
\end{pmatrix}.
\end{align}

Because $h_{\nu,2j}$ is commutable each other
($h_{\nu,2j+1}$ is also commutable each other),
it is easy to decompose $e^{{H}_{\nu,e}}$ 
and {$e^{{H}_{\nu,o}}$} as follows:
\begin{align}
e^{{H}_{\nu,e}}&=e^{{h}_{\nu,0}}\times e^{{h}_{\nu,2}}\times\cdots, \\
e^{{H}_{\nu,o}}&=e^{{h}_{\nu,1}}\times e^{{h}_{\nu,3}}\times\cdots.
\end{align}
From this relation, by just diagonalizing ${h}_{\nu,i}$ whose matrix size
is $4\times 4$, we can perform the 
real-time evolutions.

Because $H_{\nu,e}$ and $H_{\nu,o}$
are not commutable, we use 
the fourth-order Suzuki-Trotter decomposition~{\cite{Suzuki1990PhyLettA}},
whose general form is by
\begin{align}
&e^{\eta({A}_{1}+\dots+{A}_{q})}=S(\eta p)S(\eta(1-2p))S(\eta p)+O(\eta^5),\\
&S(\eta)=e^{\eta{A}_{1}/2}e^{\eta{A}_{2}/2}\cdots e^{\eta{A}_{q-1}/2}e^{\eta{A}_{q}} \notag \\
&\times e^{\eta{A}_{q-1}/2}\cdots e^{\eta{A}_{1}/2} ,\\
&p=(2-2^{1/3})^{-1},
\end{align}
where $\eta$ is c-number and 
$A_{q}$ denotes the matrix.
By using the formula,
we can decompose $e^{\eta H_{\rm W}}$
as follows:
\begin{align}
e^{\eta{H}_{\rm W}}&=S(\eta p)S(\eta(1-2p))S(\eta p),\\
S(\eta)&=S_{0}(\eta)e^{\eta{H}_{\rm diag}}S_{1}(\eta),\\
S_{0}(\eta)&=e^{\eta{H}_{x,e}/2}e^{\eta{H}_{x,o}/2}e^{\eta{H}_{y,e}/2}e^{\eta{H}_{y,o}/2} \notag \\
&\times e^{\eta{H}_{z,e}/2}e^{\eta{H}_in{z,o}/2}\notag \\
S_{1}(\eta)&=e^{\eta{H}_{z,o}/2}e^{\eta{H}_{z,e}/2}e^{\eta{H}_{y,o}/2}e^{\eta{H}_{y,e}/2} \notag \\
&\times e^{\eta{H}_{x,o}/2}e^{\eta{H}_{x,e}/2}.
\end{align}
{If the Hamiltonian is not time-dependent one,
this formula has fourth-order precision.
For the time-dependent Hamiltonian,
the time-evolution operator is defined by using the
{\it super operator} $\tilde{\mathcal{T}}$ as follows~{\cite{Suzuki1993PJA,suzuki1994}}:
\begin{align}
U(t+\Delta t,t)&=\exp{\Big[\Delta t(-{\rm i}H(t)+\tilde{\mathcal{T})}\Big]},\\
F(t)e^{\Delta t\tilde{\mathcal{T}}}G(t)&=F(t+\Delta t)G(t).
\end{align}
{Here, $F$ and $G$ are arbitrary functions. We note 
that the super operator $\tilde{\mathcal{T}}$ 
only acts on the operators on its left.}
By using this formula,
we decompose $U$ as follows:
\begin{align}
&U(t+\Delta t,t)=
S(-{\rm i}\Delta tp,t+(1-p/2)\Delta t) \notag \\
&\times S(-{\rm i}\Delta t(1-2p),t+p\Delta t/2) \notag \\
&\times S(-{\rm i}\Delta tp,t+p\Delta t/2)
+O(\Delta t^{5}),
\label{eq:ST}
\end{align}
where time-dependent $S$ is defined as
\begin{align}
S(\eta,t)&=S_{0}(\eta,t)e^{\eta{H}_{\rm diag}}S_{1}(\eta,t),\\
S_{0}(\eta,t)&=e^{\eta{H}_{x,e}/2}e^{\eta{H}_{x,o}/2}e^{\eta{H}_{y,e}(t)/2} \notag \\
&\times e^{\eta{H}_{y,o}(t)/2}e^{\eta{H}_{z,e}/2}e^{\eta{H}_{z,o}/2}\notag \\
S_{1}(\eta,t)&=e^{\eta{H}_{z,o}/2}e^{\eta{H}_{z,e}/2}e^{\eta{H}_{y,o}(t)/2} \notag \\
&\times e^{\eta{H}_{y,e}(t)/2}e^{\eta{H}_{x,o}/2}e^{\eta{H}_{x,e}/2}.
\end{align}
We note that super operator $\tilde{\mathcal{T}}$ operates
all the left side operators.
By using Eq.~(\ref{eq:ST}), we perform the 
real-time evolution.}

\begin{figure}[t!]
  \begin{center}
    \includegraphics[width=8cm,clip]{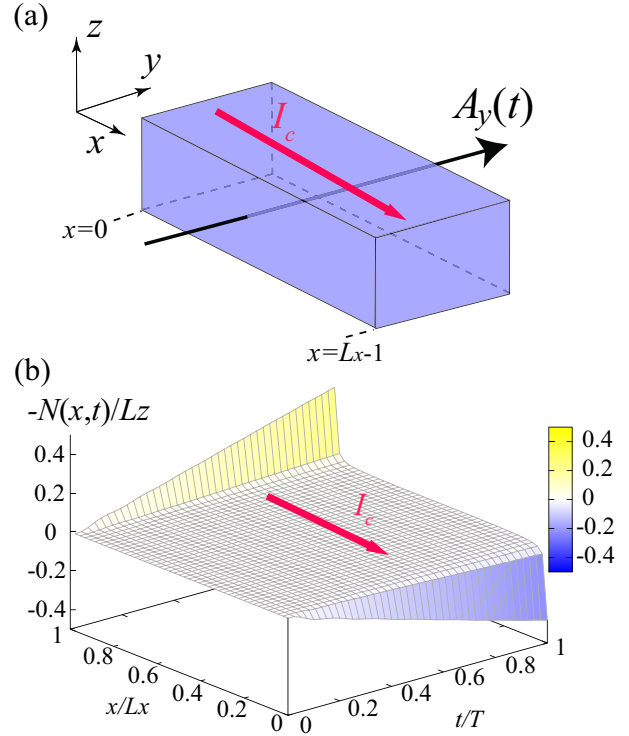}
  \end{center}
\caption{
(color online)~
(a)~Schematic illustration of the geometry used in this study.
We take open boundary condition in $x$ direction 
and the periodic boundary conditions in $y$ and $z$ direction.
Employed geometry is given by $L=L_{x}=2L_{y}=2L_{z}$.
We introduce the $A_{y}(t)$ in $y$ direction  and
charge pumping occurs in $x$ direction.
(b)~Time-dependence of the charge distribution $N(x,t)$ in $x$ direction.
We take $m=0$, $T=20$, $W=0$, and $L=44$.
The charge pumping occurs around the edges 
in $x$ direction by introducing $A_{y}(t)$.}
\label{fig:3D}
\end{figure}

\section{Results}
\subsection{Thouless pumping in the Weyl semimetals}
{In Fig.~\ref{fig:3D}(a),
we show a setup of Thouless pumping 
for the Weyl semimetal.
The system size is given by 
$N_{s}=L_{x}\times L_{y}\times L_{z}$
and we employ the rectangle geometry given by
$L=L_{x}=2L_{y}=2L_{z}$.
In $x$ ($y$ and $z$) direction, we employ the
open (periodic) boundary condition.
By applying the vector potentials in $y$
direction  {(Eq.~(\ref{eq:vecp}))}, it is expected that 
the quantized charge pumping in $x$ direction occurs.
{We take $\Delta t=0.02$ and $m=0$ in this paper}.}

\begin{figure}[h!]
  \begin{center}
    \includegraphics[width=8cm,clip]{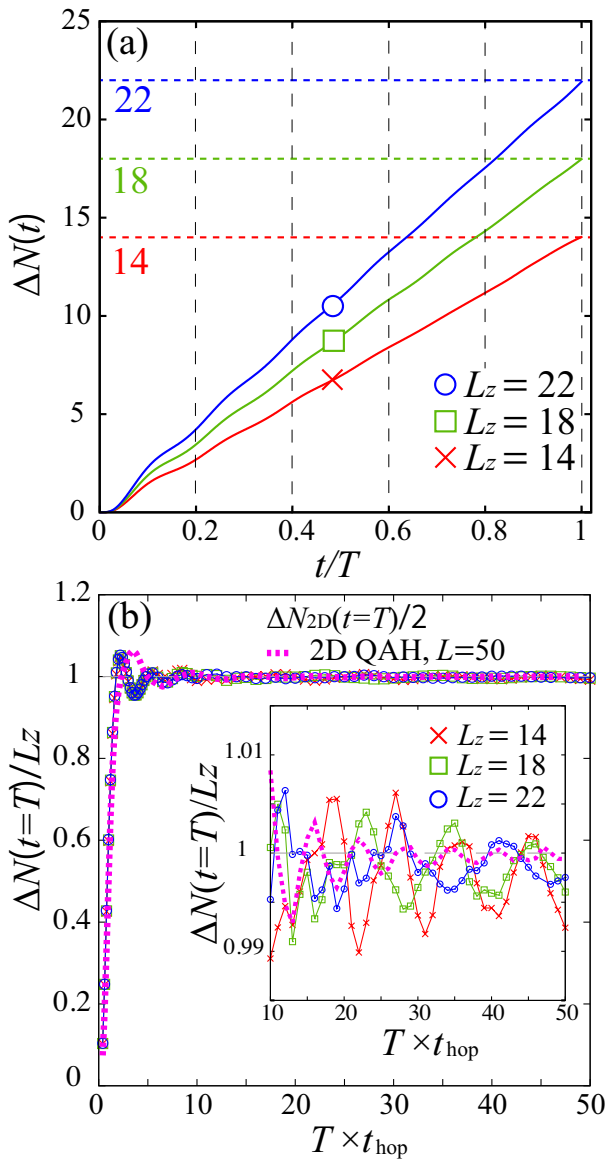}
  \end{center}
\caption{(color online)~(a)System size dependence of the charge pumping
as a function of time. We take $T=20$, 
$W=0$ and $m=0$.
The charge pumping $t=T$ is quantized as $\Delta N(t=T)=2\Delta k\frac{L_{z}}{2\pi}$,
which is proportional to $L_{z}$.
(b)~
$T$ dependence of the charge pumping.
For small $T$, because the introducing $A_{y}(t)$ is not adiabatic,
the charge pumping is not quantized. 
For comparison, we also show $T$ dependence of the charge pumping
for the two-dimensional QAH insulator $N_{\rm 2D}(t=T)$, 
where the bulk gap exists.
Details of the two-dimensional 
QAH insulator are shown in Appendix.
For $T\geq 10$,
the charge pumping is nearly quantized for both systems  
except for small oscillations. 
{In the inset, we show enlarged figure for $T\geq10$.
In contrast to the gapped system (see the inset of Fig.9 (b)),
for the Weyl semimetal, we find that
the small but finite oscillations remains even for larger $T$.
This remaining oscillations may originate from the gapless nature
of the Weyl semimetal.}
}
\label{fig:depLtime}
\end{figure}

In this setup, we perform the Thouless pumping, i.e.,
solving the time dependent Schr\"{o}dinger equations and
obtain $|\phi(t)\rangle$.
From $|\phi(t)\rangle$, we calculate 
the time-dependent charge distribution in $x$ direction, which is defined as
\begin{align}
N(x,t)=&\sum_{y,z}\Big[ \bra{\phi(t)} c_{x,y,z}^{\dagger}c_{x,y,z}\ket{\phi(t)} \notag\\
&-\bra{\phi(0)} c_{x,y,z}^{\dagger}c_{x,y,z}\ket{\phi(0)}\Big].
\end{align}
As shown in Fig.~\ref{fig:3D}(b),
by introducing the vector potentials in $y$ direction,
the charge pumping in $x$ direction occurs, i.e.,
$N(x,t)$ becomes positive around $x=0$ while
it becomes negative around $x=L_{x}-1$. 
{This result shows that the pumped charge is mainly induced
at the edges in the clean limit.}

At $t=T$, 
the total pumped charge 
is expected to be quantized for sufficiently large $T$.
Total pumped charge is defined as
\begin{align}
\Delta N(t)=\sum_{0\leq x<L_{x}/2}N(x,t)-\sum_{L_{x}/2\leq x<L_{x}}N(x,t)
\end{align}
We show $\Delta N(t)$ for several different
system sizes in Fig.~\ref{fig:depLtime} (a).
We find that
$\Delta N(t)$ monotonically increase as a function of $t$
and it nearly becomes $L_{z}$ at $t=T$. 
This is consistent with the topological properties of
the Weyl semimetals, i.e.,
the Hall conductivity is quantized as 
$\sigma_{xy}=(e^2/2\pi h)\times\Delta k$ and
the corresponding charge pumping is given by 
$\Delta N = (\Delta k/\pi)\times L_{z}$.
This result indicates that the Thouless pumping works well
even when the systems have no bulk gaps. 

To examine when the Thouless pumping can be 
regarded as the adiabatic process,
we calculate unit time ($T$) dependence of the charge pumping.
In Fig.~\ref{fig:depLtime} (b),
we show $T$ dependence of $\Delta N(t=T)$ 
for several different system sizes.
For small $T$ ($T<1$), speed of introducing $A_{y}(t)$ is 
too fast to change the electronic states in the Weyl semimetals.
Thus, for $T<1$, the Thouless pumping is non-adiabatic and 
the pumped charge is not quantized. 
By increasing $T$, for {$T\geq 10$},
the pumped charge is quantized except for small oscillations.
This result indicates that the Thouless pumping can be regarded 
as the adiabatic process for $T\geq10$.
Thus, we take $T=20$ in the most {remaining} part of this paper.
{
We note that the typical time scale does not significantly change for weak disorder region 
but it becomes large for strong disorder region. 
Nevertheless, we note that the charge pumping at $T=20$ 
can be regarded as the adiabatic pumping in the relevant disorder region.}

We note that the Laughlin's argument or the Thouless's argument
requires the existence of the bulk charge gap for
the quantized charge pumping.
The Weyl semimetal has no bulk charge gap and it is unclear
whether the Thouless pumping works well or not.
By comparing with $T$ dependence in the two-dimensional QAH insulator
as shown in Fig.~\ref{fig:depLtime} (b),
we find that $T$ dependence of the 
pumped charge for the Weyl semimetal is basically 
the same as that of the QAH insulator.
This result clearly shows that the Thouless pumping
works well for detecting the topological 
invariant even when the systems have no bulk gap.

\subsection{Effects of disorders}
We examine how the disorder affects
the charge pumping in the Weyl semimetal.
In general, topological property is robust 
against the perturbations
because the topological property 
can not be changed by the perturbations
unless the energy scale of the perturbations reached that of the charge gap.
For the Weyl semimetal, it is, however, unclear
whether topological property remains or not 
because the bulk charge gap is zero in the Weyl semimetal.
Several theoretical studies, however, show that
topological properties in the Weyl semimetals are robust 
against the small disorder~\cite{Chen_PRL2015,Shapourian_PRB2016,Takane_JPSJ2016,Liu_PRL2016}. 
We examine whether the Thouless pumping can reproduce the results of
the previous studies.
{We note that
we do not consider the rare region 
effects~\cite{Nandkishore2014PRB,Pixley2016PRX,Lee2018PRB} in this paper.}

In Fig.~\ref{fig:modt1}, we show the disorder dependence of 
total charge pumping $\Delta N(t=T)$.
We find that overall behaviors are consistent 
with previous studies~\cite{Chen_PRL2015,Liu_PRL2016,Shapourian_PRB2016,Takane_JPSJ2016};
plateau for small disorder {($W\lesssim1$)}, 
enhanced charge pumping in
the intermediate disorder ($W\sim3$), 
and decrease of the charge pumping
in the strong disorder region {($W\gtrsim4.5$)}.
We note that enhancement of the 
charge pumping is not observed in the two-dimensional
QAH insulator as shown %in Fig.~\ref{fig:2D} 
in Appendix.

\begin{figure}[tb!]
  \begin{center}
    \includegraphics[width=7cm,clip]{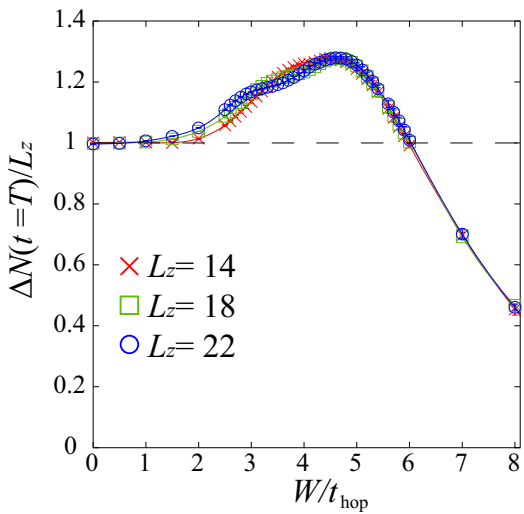}
  \end{center}
\caption{(color online)~
Disorder dependence of 
the pumped charge  
at $t=T$.
We take $T=20$ and $m=0$.} 
\label{fig:modt1}
\end{figure}

To examine how the pumped charge is enhanced by the disorder,
we analyze the real-space dependence of the pumped charge.
In Fig.~\ref{fig:W}, 
we show the charge distribution at $t=T$ for 
several different strengths of the disorders.
Because the charge pumping mainly occurs around the
edges ($x\sim 0,L_{x}-1$),
we enlarge the shaded region in Fig.~\ref{fig:W}(a) and
plot the $x$ dependence of the pumped charge
measured from the clean limit ($\Delta\tilde{N}(x,t)=\Delta N(x,t,W)-\Delta N(x,t,W=0)$)
in Fig.~\ref{fig:W}(b). 

For small disorder ($W=2$),
we find that the disorder mainly changes the pumping 
around the edges and it does not affect the pumping 
inside of the systems.
{This enhancement for small disorder 
can be explained by the mass
renormalization effects~\cite{Chen_PRL2015,Liu_PRL2016,Shapourian_PRB2016,Takane_JPSJ2016}, 
i.e. disorder increases mass term $m$ 
and widen the length of Fermi arcs.}
By further increasing the strengths of the disorders,
we find that the pumped charge begins 
to penetrate into the systems.
This behavior can be explained as follows:
For the strong disorder region,
{the Fermi arcs at the surfaces} 
begin to mix with the bulk states.
This mixing induces the penetration of the
Fermi arcs inside the systems, i.e., Fermi arcs begins to have finite
width in $x$ direction and induces the charge pumping
inside of the systems.
This is the reason why the pumped charge is enhanced by the disorder.

\begin{figure}[t!]
  \begin{center}
    \includegraphics[width=7cm,clip]{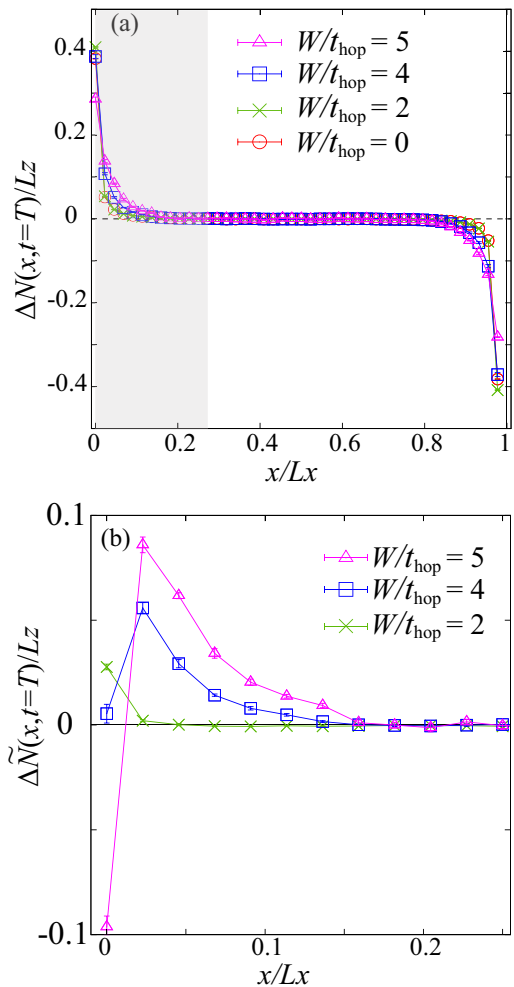}
  \end{center}
\caption{(color online)~(a)~Charge pumping in the presence of the disorder.
We plot $\Delta N(x,t)/L_{z}$ at $t=T$ for several different strengths of 
disorder. To estimate the errors of the realizations of the 
disorder, we take five different realizations and regard 
its standard errors as error bars.
We take $L=44$ and  $m=0$. 
(b) Pumped charge measured from the clean limit 
[$\Delta \tilde{N}(x,t=T)$]
in the shaded region in (a).
By increasing the strength of the disorder,
we find that the pumped charges penetrate into the inside of the 
system. 
}
\label{fig:W}
\end{figure}

\begin{figure}[b!]
  \begin{center}
    \includegraphics[width=7cm,clip]{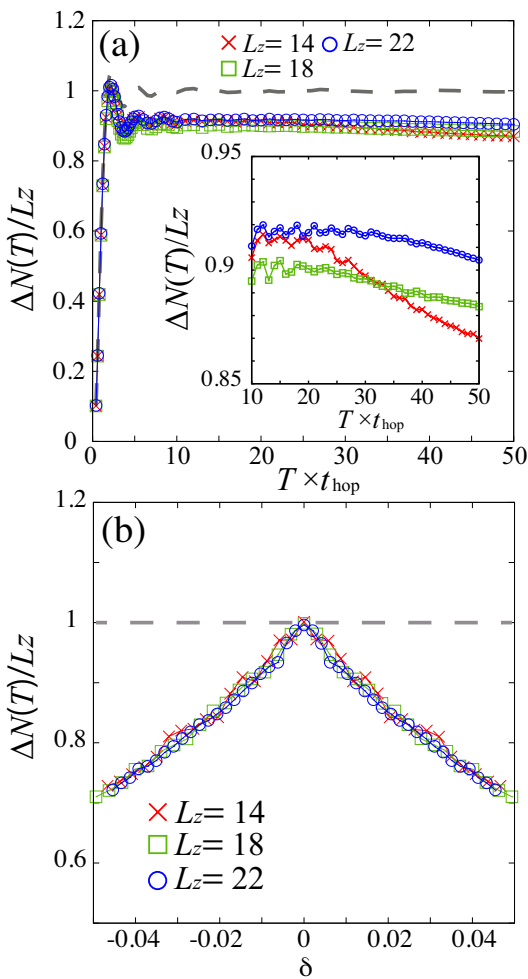}
  \end{center}
\caption{(color online)~(a)~
$T$ dependence of 
the charge pumping $\Delta N(t=T)/L_{z}$ at finite doping $\delta = 0.01$.
{We take $m=0$.}
For comparison, we plot result at zero doping by broken lines.
{In the inset, we show enlarged figure for $T\geq 10$.}
%In the larger $T$ region, the pumped charges slightly decrease as a function of $T$.
%However, the origin of the small decrease may be the finite-size effects because 
%%the pumped charge at fixed $T$ increases 
%by increasing $L_{z}$ for $T\geq 30$ and it is expected to
%converge to value for $T\leq 20$.
%Thus, we can roughly regard the values at $T=20$ 
%as the converged value in the long-time
%limit and the bulk limit.}
(b)Doping dependence of charge pumping for $T=20$.
Charge pumping monotonically decrease by doping.}
\label{fig:Tdope}
\end{figure}

\subsection{Effects of doping}
In this subsection, we examine the effects of doping
on the charge pumping 
in the Weyl semimetal. 
First, we examine whether the
adiabatic limit of the
charge pumping exists for finite doping 
where the finite density of states exists.
In Fig.~\ref{fig:Tdope}(a),
we show $T$ dependence of the pumped charge.
At small $T$ ($T\leq 1$), i.e., at the non-adiabatic process,
$T$ dependence of the pumped charge is same as that of zero doping.
{Here, the doping rate is defined as 
$\delta=\Delta N/N_{s}$, where
$\Delta N$ is the number of electrons measured from 
half filling, i.e., $\Delta N=N_{e}-N_{s}$.}
In the non-adiabatic region, 
because the electrons move too fast,
low-energy structures of the systems
such as the Fermi surfaces do not affect the 
charge pumping.
This is the reason why the pumped charges
do not change in the non-adiabatic region.

By taking larger $T$, 
{we find that 
$T$ dependence of the charge pumping is basically
the same as that of the non-doping case, i.e.,
the large oscillations seen for small $T$ ($T\leq 5$) are suppressed 
and the pumped charge seems to converge to the constant for $T\leq20$. 
However, as shown in the inset in Fig.~\ref{fig:Tdope}(a), 
it slightly decreases for $T\geq20$ and 
there is considerable system-size dependence.
From the available data,
it is difficult to 
identify whether the origin of the decrease 
is the finite-size effects or not 
and it is also difficult to 
accurately estimate the converged pumped charge 
in the long-time and bulk limit for the doped case. 
Nevertheless, as we show later,
the pumped charge around $T=20$ can capture the
essence of the finite doping effects on the charge pumping and
can be useful for detecting the remnant 
of the quantized charge pumping at
zero doping.
Thus, to examine the doping effects,
we use the pumped charge at $T=20$ 
as a simple estimation of the converged value.
}
%This result indicates that 
%the adiabatic limit can be defined even for finite doping.
%{In other words, the Thouless pumping can detect 
%the remnant of the quantized charge pumping.}

In Fig.~\ref{fig:Tdope}(b),
we show doping dependence of the pumped charge for $T=20$.
We find that the pumped charge monotonically decreases for electron and hole doping
except for slight oscillations found in small system sizes.
This result shows that 
doping into the Weyl semimetals 
continuously lowers
pumped charge from its quantized values at zero doping.
{We note that the changes in the pumped charge are
induced by the Berry curvature in non-linear dispersions
because the Berry curvature in linear dispersions around the
Weyl points does not contribute to the pumped charge.}
{We note that the saddle points around the zero doping are located at 
$E_{\rm saddle}=\pm t_{\rm hop}$ and corresponding doping rate is given by
$\delta_{\rm saddle} \sim \pm 0.065.$}

We examine the disorder 
effects of Thouless pumping 
at finite dopings.
In Fig.~\ref{fig:dopeW},
we plot disorder $W$ dependence of the 
pumped charge for several different doping rates. 
At finite doping rates, 
{quantization} at zero doping is absent,
the pumped charge monotonically {increases} as a function
{the disorder strength $W$.}
By further increasing the disorder, the pumped charge 
has peaks around $W\sim 4.5$ as in the case of the zero doping.
The robustness of the peak structure against the finite doping rates
is a 
{characteristic} feature of the Weyl semimetal.
In the strong disorder region ($W\geq6$), charge pumping
does not depend on the doping rates because the Fermi surfaces
are completely smeared out in this region.

\begin{figure}[h!]
  \begin{center}
    \includegraphics[width=7cm,clip]{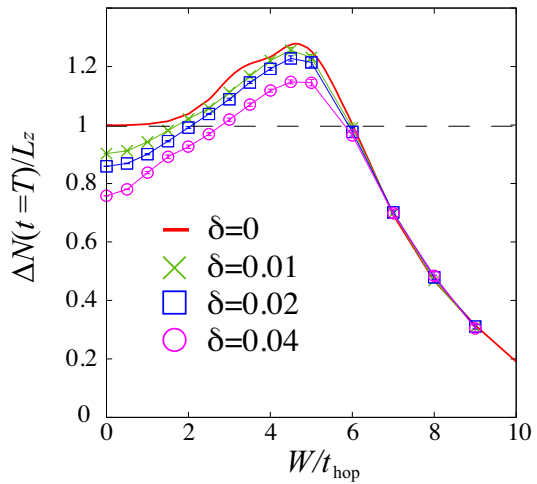}
  \end{center}
\caption{(color online)~Disorder $W$ dependence of 
pumped charge at finite doping. We take $T=20$ and $m=0$.}
\label{fig:dopeW}
\end{figure}

\section{Summary}
To summarize,
we have introduced the lattice model for the Weyl semimetal in Sec.2.A and
have detailed the methods for solving 
the time-dependent Schr\"{o}dinger equations by using the 
fourth-order Suzuki-Trotter decomposition in Sec.2.B.
Although the time-dependent Schr\"{o}dinger equations
can be solved by performing the diagonalization
of the Hamiltonian at each time step,
numerical cost of diagonalization is large 
and that method can not be applied to the large systems. 
The Suzuki-Trotter decomposition method does not require
the diagonalization of the full Hamiltonian at each time step, 
numerical cost is dramatically reduced. 
By using this method, we can perform 
the Thouless pumping up to the order of $10^4$-sites systems.

In Sec. 3.A, we have shown the results of the Thouless 
pumping for clean limit and zero doping.
Although the Weyl semimetal does not have the 
bulk charge gap, we have found that the Thouless 
pumping works well for detecting the topological quantization
of the Weyl semimetals.
By examining the unit time $T$ dependence of the 
charge pumping, we confirm that {the} adiabatic charge pumping
occurs for larger $T$, typically $T\geq20$.

In Sec. 3.B, we have examined the disorder effects on 
the Thouless pumping. We note that the Thouless pumping
itself can be applied to the disorder systems
without changing the method because we just solve
the time-dependent Schr\"{o}dinger equations in the real space.
As a result, we have shown that the quantized pumped charge 
is robust against small disorder.
We have also shown that the pumped charge increases by 
increasing the disorder {for the intermediate strength of disorder.}
These behaviors are consistent with the 
previous studies~\cite{Chen_PRL2015,Liu_PRL2016,Shapourian_PRB2016,Takane_JPSJ2016}.
This 
shows that the usefulness of the Thouless pumping
for detecting the topological properties in the disordered systems.

We have found that the charge pumping has 
large system size dependence around $W\sim 3$ as shown in Fig.\ref{fig:modt1},
where the transition between Weyl semimetals
and diffusive metal is 
pointed out in the literature~\cite{Shapourian_PRB2016}.
Thus, this system size dependence may be related to 
the transition into the diffusive metal.
In this study, available system size is limited and
it is difficult to perform the accurate finite-size scaling
for detecting the signatures of the phase transitions.
Systematic calculations for determining the phase transitions
is left for future studies.

In Sec. 3.C, we have examined the effects of the 
doping into the Weyl semimetals.
%Even at finite doping rates,
%we have found that the pumped charges converge to
%the constant values for large unit time $T$, i.e.,
%adiabatic pumping also occurs at finite doping rates.
{For finite doping rates, we have found that
the pumped charge slightly decreases for larger $T$
and it is difficult to 
accurately estimate the pumped charge
in the adiabatic limit.
In this paper, we simply use the pumped charge at $T=20$
as a rough estimation of the adiabatic pumping.
It is left for future studies to accurately estimate
the pumped charge in the adiabatic limit
by performing calculations for 
larger system sizes and larger $T$.}
{By using the pumped charge at $T=20$,
we have shown that the remnant of the quantized pumped charge 
can be detected for finite doping rates.}
We have also shown that the pumped charge is also enhanced 
by increasing the disorders for finite doping rates. 
{The} peak positions of the charge pumping under {disorder}
do not largely depend on the doping rates.

Our results show that the Thouless pumping is a
useful theoretical tool for detecting the 
topological properties even for the gapless 
systems such as the Weyl semimetals.
This method is also applicable to the 
doped systems and can capture remnant of the 
topological properties of the systems 
through the charge pumping.
Because the Thouless pumping only requires the 
real-time evolution of the ground-state wave functions,
it can be applied to the correlated electron systems where
it is difficult to obtain the full eigenvectors.
For the one-dimensional system,
the Thouless pumping for the correlated system is
studied in detail~\cite{Nakagawa2018PRB}. 
Recent studies~\cite{Haegeman2011,Carleo2012,Ido_PRB2015} show that it is possible to
perform the accurate real-time evolutions of the wavefunctions
in the correlated quantum many-body systems 
based on the time-dependent variations principles~\cite{Mclachlan_MP1964}. 
Studies in this direction are intriguing challenges
for clarifying the nature of the correlated topological systems 
{in more than one dimension} 
and our detailed study on the Thouless pumping presented in this paper
offers a firm basis for such advanced studies.

\begin{acknowledgements}
Our calculation was partly carried out at the 
Supercomputer Center, Institute for Solid State Physics, University of Tokyo.
This work was supported by JSPS KAKENHI 
(Grant Nos.~JP15H05854,~JP16H06345,~JP16K17746,
~{JP17K05485},~{JP17K17604},~{JP19K03739}).
TM thanks Ken-Ichiro Imura for useful discussions on real-time evolutions.  
{We also thank Koji Kobayashi for useful discussions on 
the disordered magnetic Weyl semimetal.}
TM was also supported by Building of Consortia for 
the Development of Human Resources in Science and Technology from the MEXT of Japan.
{This work was also supported by 
the Japan Society for the Promotion of Science, and JST
CREST (JPMJCR18T2).}
\end{acknowledgements}

%\bibliographystyle{apsrev}
%\bibliography{topo}

\appendix
\clearpage
\section{Thouless pumping in two dimensional Chern insulators}
Here, we show the results of the Thouless pumping 
for the two-dimensional quantum anomalous Hall (QAH) insulators.
By simply ignoring the $z$ dependence of the Weyl Hamiltonians,
we can obtain the lattice Hamiltonian for the
two-dimensional QAH insulators as follows:
\begin{align}
&H_{\rm QAH}=\sum_{\nu=x,y}H_{\nu}+H_{\rm diag},\\
&H_{\rm diag}=(2-m)\sum_{j}c_{j}^{\dagger}\sigma_{z}c_{j}+\sum_{j}\epsilon_{j}c_{j}^{\dagger}\sigma_{0}c_{j},\\
&\epsilon_{j}\in [-W/2,W/2].
\label{eq:Weyl}
\end{align}
For $m>0$, we obtain the QAH insulator with $\mathcal{C}=1$ and
trivial insulator appears for $m<0$.
We consider $L=L_{x}=L_{y}$ systems and
the pumped charge is given by
\begin{align}
\Delta N_{\rm 2D}(t)=\sum_{0\leq x<L_{x}/2,y}N(x,y,t)-\sum_{L_{x}/2\leq x< L_{x},y}N(x,y,t)
\end{align}
We note that the charge pumping is quantized as follows:
\begin{align}
\Delta N_{\rm 2D}(t=T) =2\times \mathcal{C},
\end{align}
where $\mathcal{C}$ is the Chern number.

\begin{figure}[tb!]
  \begin{center}
    \includegraphics[width=7cm,clip]{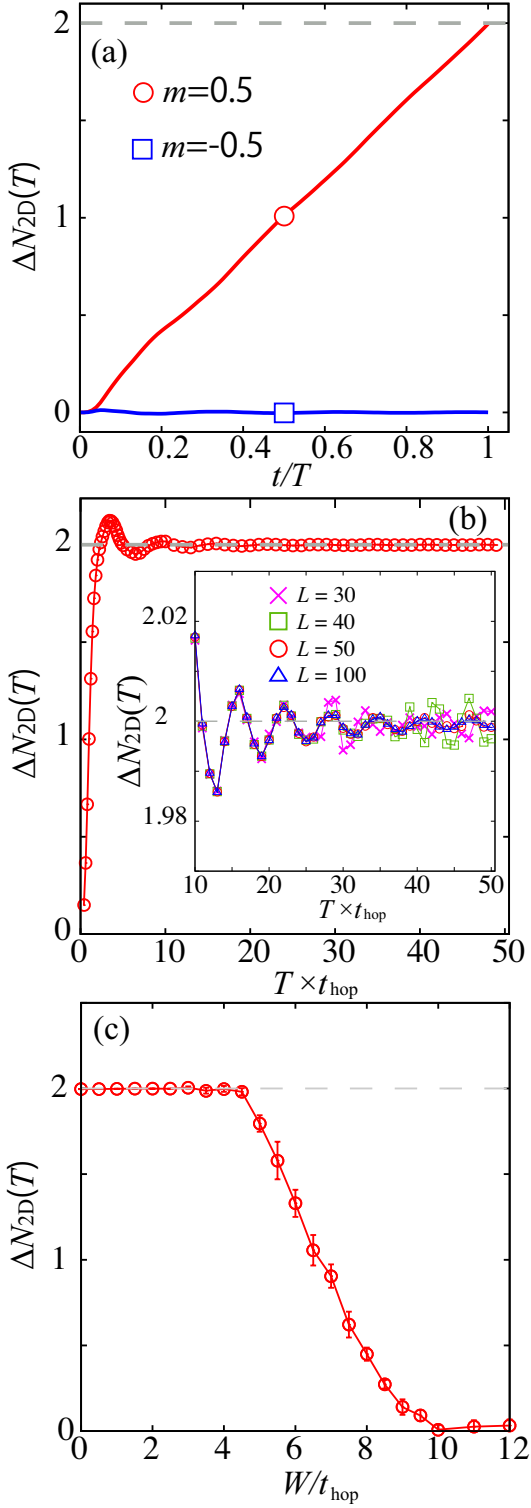}
  \end{center}
\caption{(color online)~
(a) Thouless pumping for the Chern insulator 
($m=0.5$) and trivial insulator ($m=-0.5$).
We take $L=50$ and $T=20$.
(b)~
$T$ dependence of charge pumping
for $L=50$ and $m=0.5$.
{In the inset, we show the enlarged figure for $T\geq 10$.
In the QAH systems, we find that the oscillations in the pumped charge
become smaller for larger $T$.}
(c)~Disorder dependence of the 
pumped charge.}
\label{fig:2D}
\end{figure}

In Fig.~\ref{fig:2D}(a), we show the results of Thouless pumping
for $m=0.5$ and $m=-0.5$. In the {topologically} trivial insulator ($m=-0.5$),
the charge pumping does not occur while the charge pumping
is quantized for $m=0.5$. Because the 
{Chern} number is 1 in this system,
the quantized charge pumping becomes 2.

We show $T$-dependence of the $\Delta N_{\rm 2D}(T)$ in Fig.\ref{fig:2D}(b).
Similar to the Weyl semimetals, 
although the oscillation occurs for small $T$ ($T\leq 10$),
the charge pumping converges to the quantized value. 
In the QAH insulators, 
we only show the results for $L=50$ 
because size effects are small.

We show the disorder dependence of the $\Delta N_{\rm 2D}(T)$
in Fig.\ref{fig:2D}(c). In contrast to the Weyl semimetals,
the charge pumping does not have peak structures.
For $W\leq5$, the pumped charge is 
quantized 
and it begins to decrease for $W\geq 5$. 
This result indicates that the characteristic enhanced
charge pumping in the Weyl semimetal is induced by
its gapless nature.

\end{document}